\documentclass[aps,twocolumn,floatfix,showpacs]{revtex4}
\usepackage{graphics}
\newcommand{\la}{\lambda}
\newcommand{\bea}{\begin{eqnarray}}
\newcommand{\beq}{\begin{equation}}
\newcommand{\eea}{\end{eqnarray}}
\newcommand{\eeq}{\end{equation}}
\def\simleq{\; \raisebox{-0.4ex}{\tiny$\stackrel
{{\textstyle<}}{\sim}$}\;}
\begin{document}
\title{Resonance scattering and singularities of the scattering function}
\author{W.\ D.\ Heiss$^{1}$ and R.\ G.\ Nazmitdinov$^{2,3}$}         
\affiliation{$^{1}$National Institute for Theoretical Physics, \\
Stellenbosch Institute for Advanced Study, \\ 
and Institute of Theoretical Physics,
University of Stellenbosch, 7602 Matieland, South Africa\\
$^{2}$Department de F{\'\i}sica,
Universitat de les Illes Balears, E-07122 Palma de Mallorca, Spain\\
$^{3}$ Bogoliubov Laboratory of Theoretical Physics,
Joint Institute for Nuclear Research, 141980 Dubna, Russia}

\begin{abstract}
Recent studies of transport phenomena with complex potentials
are explained by generic square root singularities of spectrum and eigenfunctions of
non-Hermitian Hamiltonians. Using a two channel problem we
demonstrate that such singularities produce a significant effect upon
the pole behaviour of the scattering matrix, and more significantly upon the
associated residues. This mechanism explains why by proper choice
of the system parameters the resonance cross section is increased drastically
in one channel and suppressed in the other channel.
\end{abstract}
\maketitle
\section{Introduction}
\label{intro}
It is well established knowledge that the shape of measured cross
sections, in particular that of resonances, is described
satisfactorily by poles of the scattering function in the complex
energy plane \cite{newton}. These poles correspond to particular
solutions of the Schr\"odinger equation at complex energies by
imposing the boundary conditions that the wave function be regular at
the origin and have only outgoing waves at large distances \cite{mahaux}.
They have been discussed recently again in the context of complex
potentials \cite{mosta}. In fact, while for a hermitian Hamiltonian such
solutions can occur only at complex energies with non-vanishing
negative imaginary part \cite{foot}, a complex potential (associated with
absorption)
can give rise to such states at energies with vanishing or even positive imaginary part.
As a rule, these specific solutions give rise to a pole of first order
in the Green's function $(E-H)^{-1}$ and thus in the scattering
function. This remains true also in multi-channel problems, even if
there is a degeneracy, where the residue of the pole is a matrix of
rank larger than one.

There are, however, special singularities of the Green's function and thus of
the scattering function being of a very different nature: the
exceptional points (EP) of a Hamiltonian. They are also specific solutions
of the Schr\"odinger equation or, in more general terms, of an
eigenvalue problem, but for a non-hermitian problem. They have been discussed
at great length theoretically \cite{he,sey,gu} and in a large variety of
applications such as in atomic physics \cite{sol,cart}, in optics \cite{ber1},
in nuclear physics \cite{ok} and in different theoretical
context in ${\cal PT}$-symmetric models \cite{zno}, to name just a few.
Depending on the particular situation EPs can signal a phase transition \cite{hege,cej}.
The experimental verification of the existence of EPs
requires a careful tuning of parameters of the particular open system investigated
(see below Sect.\ref{sec:2}).
Recently the topological structure of the square root branch point associated
with an EP has been shown experimentally to be a physical reality \cite{demb1,demb2,ep}.

While some of the quoted papers report about particular physical
effects of the EPs, either expected or measured, no emphasis is
placed upon the important role of the eigenstates when the vicinity
of an EP is explored in the laboratory. We recall that an EP is
characterised by the coalescence of not just two eigenvalues but also of
their associated eigenstates; moreover, the norm of the associated
eigenstate - there is only one - vanishes at the EP. It is the main
purpose of the present paper to illustrate the dramatic effect of this
vanishing norm upon the residues of the scattering function (which
has a pole of second order at the EP \cite{mond}) resulting
in a distinctly different resonance behaviour in the different channels. To
facilitate the demonstration we restrict ourselves to a two-channel
problem.

\section{Sharp resonances and Exceptional Points}
\label{sec:1}
\subsection{The Model}
\label{sec:1a}
To illuminate the basic mechanism  of the sudden increase of a resonance cross
section, associated with EPs, we strip the great variety of models down to the essentials and
ela\-borate the universal cause by studying  a generic model.
It is well established that, in the close vicinity of an EP, a
two-dimensional matrix model suffices to capture all essential
features associated with the singularity. We thus begin with
the model Hamiltonian
\begin{eqnarray}
H(\la )&=& H_0+H_1(\la )=H_0+\la V \nonumber\\
&=&
\left(
\begin{array}{cc}
\omega_1 & 0 \\
0 & \omega_2
\end{array}
\right)
+\la
\left(
\begin{array}{cc}
\epsilon_1 & \delta \\
\delta & \epsilon_2
\end{array}
\right)
\label{ham}
\end{eqnarray}
where the parameters $\omega_k$ and $\epsilon_k$ determine the non-in\-ter\-ac\-ting resonance
energies $E_k=\omega_k+\la \epsilon_k, \,k=1,2$. They are chosen complex and
such that the $E_k$ have negative imaginary parts;
we focus on $\la $ between 0.5 and 0.6. As we are interested in
the effects of crossing and coalescence in scattering, we
consider in the following the scattering matrix, {\it viz.}
\begin{equation}
T_{i,k}=H_1(\la )_{i,k}+(H_1(\la )\,(E-H(\la ))^{-1}\,H_1(\la ))_{i,k}.
\label{scat}
\end{equation}
Since we mimic a multi-channel problem by an effective
two-channel Hamiltonian, i.e.~$\omega_i$ and $\epsilon_i$
are complex, the scattering amplitude (\ref{scat}) does
not satisfy the unitarity condition \cite{newton}.

To describe, for $\delta =0$, the simultaneous observation of the two
resonances in both channels on equal footing,
we have to consider the scattering
matrix in a basis rotated by the angle
$\pm \pi/4$ rather than in the basis given by (\ref{ham}).
That is, the labels $i,k$ in the scattering matrix (\ref{scat})
refer to the basis
$\{c1/\sqrt{2}\pm c2/\sqrt{2},c1/\sqrt{2}\mp c2/\sqrt{2}\}$ with
$\{c1,c2\}$ being an eigenvector of (\ref{ham}).
This specific observational basis is essential to ensure that even for the
non-interacting case ($\delta =0$) both resonances are equally and
simultaneously present in either channel; the difference between the
two channels shows when $\delta $ is switched on as is disccussed in
the following sections.
We choose the parameters $\omega_k$ and $\epsilon_k$
for convenient demonstration, that is
$\omega_1=1.55-0.007i,\,\omega_2=1.1-0.007i,\,
\epsilon_1=-0.4-0.0006i,\,\epsilon_2=0.4+0.0005i$.
The values have no particular significance, they serve to illustrate the
principle, that is the effect of a near EP upon scattering; any other
set that invokes crossing and coalescence of resonances would serve
the same purpose as long as the imaginary parts of the poles are near to
the real axis.

\subsection{The interacting resonances}
\label{sec:2}

\begin{figure}
\resizebox{0.45\textwidth}{!}{%
\includegraphics{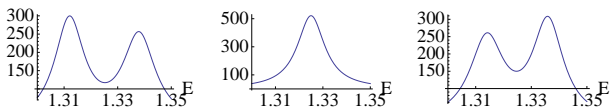}
}
\caption{Crossing of resonances without coupling ($\delta =0$).
The three figures are taken for
$\la =0.53, \Re \la _{cross}=0.5625$ and 0.59, respectively. The units are arbitrary
but the same for all three figures.}
\label{cross}
\end{figure}

From (\ref{scat}) we obtain the cross sections $\sim (TT^{\dagger})_{kk}$
as a function of the energy for various values of $\la $. This is
illustrated in Fig.1, where $(TT^{\dagger})_{11}$ is plotted for $\delta =0$.
Without coupling, when the energies cross (a usual degeneracy), the
figures for $(TT^{\dagger})_{22}$ coincide with those of
$(TT^{\dagger})_{11}$. Such crossing of resonances
is often associated with a particular symmetry,
this aspect is of no significance for our purpose.
The crossing happens at $\la_{\rm cross} =0.5625-0.00077i$.

The difference between $(TT^{\dagger})_{11}$ and $(TT^{\dagger})_{22}$
changes dramatically when the coupling $\delta $ is turned on.
In our example we have chosen a purely absorptive coupling as then the
two EPs avoid the real $\la $-axis (recall that the Hamiltonian
is non-Hermitian even for $\delta =0$ owing to the unperturbed
widths; it means EPs may occur for real $\la $).
For $\delta=0.0115 i$ the EPs are at $\la _+=0.579 - 0.00082 i$
and $\la _-=0.547 - 0.00073i$ with the energies $1.325 - 0.007029i$ and
$1.325 - 0.007027i$, respectively. The EPs sprout from $\la_{\rm cross}$,
a critical value used below is $\la _{cr}=\Re (\la _+ +\la_-)/2\approx\la_{\rm cross} $.
The dependence on any parameter is highly
sensitive when $\delta \ne 0$ owing to the proximity
of the EPs invoking large derivatives of spectrum
and eigenfunctions.
In particular, the dependence on $\delta $ is crucial and can change
patterns as is further discussed below.

\begin{figure}
\resizebox{0.35\textwidth}{!}{%
\includegraphics{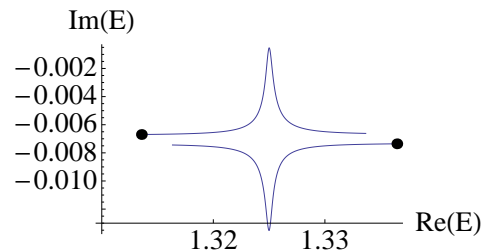}
}
\caption{Energy eigenvalue trajectories when $\la $ sweeps from 0.53 to 0.59
for $\delta =0.0115 i$. If $\delta =0$ the straight lines continue
as straight lines avoiding the peaks. Note that for our choice of parameters the
trajectories run in opposite directions, the respective starting
points are indicated by a dot.}
\label{traj}
\end{figure}

In Fig.2 we illustrate the trajectories of eigenvalues of the
Hamiltonian (\ref{ham}) obtained by
variation of $\la $ for the parameter set considered above.
We note the sharp deviations
from the straight lines of the trajectories bringing one resonance
close to the real energy axis while the other is moving into the
opposite direction. The two EPs lie basically within
the centre of the two peaks; at each of
these points the levels behave to lowest order as
$E_{\pm}\sim E_{EP}\pm {\rm const.} \sqrt{\la -\la _{EP}}$
implying a large derivative.
Moreover, the state vectors have a similar strong dependence:
while they are properly normalised their directions change rather swiftly when
$\la $ is sweeping over the critical value $\la  _{cr}=0.563$.
We emphasise that the
phenomenon of width repulsion associated with level crossings
as illustrated in Fig.2 is usually not encountered for real values
of the strength parameter $\lambda $. It occurs in our model as both
EPs are on the same side of the real $\lambda $-axis being a
consequence of dealing with a non-Hermitian Hamiltonian from the
outset \cite{heiss}. The combined effect is the
dramatic change for both cross sections in the immediate vicinity
of the critical point as illustrated in Fig.3.

The obvious effect is the huge increase of the cross section for
$(TT^{\dagger })_{11}$ which is due to the narrow resonance. In
contrast, there is a substantial drop for $(TT^{\dagger })_{22}$ in
comparison with the centre figure of Fig.1.
Moreover, the shape of this latter 'resonance' is no longer of the usual
Lorentzian type (as was first noticed in Ref.\cite{raikh}).
This dramatic change essentially takes place, albeit
continuously, within the narrow interval $\{\la  _{cr}-0.03,\la_{cr}+0.03\}$.
In fact, for the same values of the coupling the curves
at $\la =0.53$ or $\la =0.59$ look just like those in Fig.1, where
$\delta =0$ is considered. This is understandable as the trajectories
shown in Fig.2 are virtually unaffected by the small coupling
outside the interval $\{\la  _{cr}-0.03,\la  _{cr}+0.03\}$. The same holds for
the eigenvectors.
We note that the ``peak on the peak'' on the
right hand side of Fig.3 is again an effect of fine tuning. It is
brought about by the conspiracy of the two poles and their respective
residues as they occur on the top and bottom cusp in Fig.2. Decreasing
for instance $\delta $, the ``peak on the peak'' can disappear but the
resulting shape still is not a Lorentzian, it rather retains the flat
plateau without the ``second peak''.

\begin{figure}
\resizebox{0.51\textwidth}{!}{%
\includegraphics{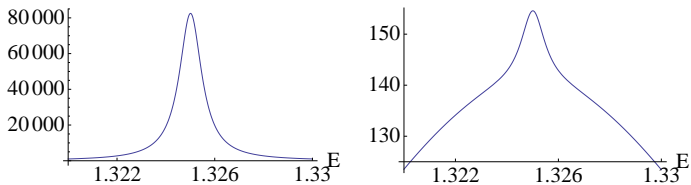}
}
\vspace*{0.32cm}
\caption{Energy dependence of $(TT^{\dagger })_{11}$ (left) and
$(TT^{\dagger })_{22}$ (right) for $\delta =0.0115i$ at the critical
point $\la  _{cr}=0.563$. Note the huge
difference in scale between the two drawings. The units
are as in Fig.1.}
\label{crit}
\end{figure}

Of course, there is nothing special about the 11-channel that becomes
so strongly enhanced. When the sign of the rotational angle $\pi /4$ is
changed, the two eigenfunctions simply swap and it is then the 22-channel
that appears so dramatically enhanced while the 11-channel is
suppressed; similarly, as the phases of the (normalised) eigenvectors of (\ref{ham})
are arbitrary, changing the sign of either $c1$ or $c2$ will also swap the roles
of the two channels. The important point is the strong enhancement of one channel
while the other is suppressed. In either case, the
effect occurs only in the immediate vicinity of $\la _{cr} =0.563$.
The pattern in Fig.3 has its origin not only in
the square root branch points of the spectrum but rather in the associated
eigenstates. This aspect, that is the effect
of the eigenstates upon the residues of the Green's function, is often
overlooked in the literature; {\it it is not only the pole positions
but also the strong and distinctly different variation
of the two residues} that brings about the
effects discussed here.

The trajectories of the spectrum in the energy plane
as illustrated in Fig.2 provide the major clue
for an understanding. Within the range of interest for
real $\la $-values the two peaks pointing in opposite directions
are well described by
\beq
E_{\pm}(\la )\approx E_r\pm \sqrt{(\la -\la _{-})(\la -\la _{+})},
\label{cxen}
\eeq
where
$$\la _{\pm}=\mp\frac{i(\omega _1-\omega _2)}{2\delta \pm i(\epsilon _1-\epsilon _2)}$$
are the two EPs. For small $\delta $ they lie near to each other
and degenerate into a single diabolic point when the coupling parameter
$\delta $ vanishes. The constant $E_r$ ensures that both
energies have a negative imaginary part for real $\la $ to allow their
interpretation as true resonances. Notice that care has to be taken
for the values to remain in the same Riemann sheet in a
numerical generation of a plot like the one in Fig.2
where real values of $\la $ are used.
At the critical point $\la _{cr}$  -- that is where the two tips
of the two peaks occur -- one energy value
is very close to the real axis while the other
is more remote from the real energy axis. These two pole terms
occur in the Green's function, and it is here where the corresponding
residues play a decisive role.
The analytical expressions for the residues blow at the EPs like
$1/\sqrt{(\la -\la _{+})(\la -\la _{-})}$ owing to the
vanishing norms of the eigenstates. We give
a simplified expression representing numerically the correct analytical
terms quite well for real $\la $. It reads
\begin{equation}
r_{\pm}(\la ) \approx \frac{1}{2}\pm\frac{A}{\sqrt{\la -(\la _{+}+\la _{-})/2}}
\label{res}
\end{equation}
where the constant $A$ is suitably adjusted yielding \newline 
$|r_{+}(\la )|\simleq 1$ at $\la _{cr} $. A combined diagram of the
real functions $|r_{+}(\la )|$ and $|r_{-}(\la )|$,
as a function of $\Re \la $,
looks similar in shape to Fig.2 with the difference
that the peaks reach near unity and zero, respectively.
The diagonal elements of
the full Green's function are then given by
\bea
(E-H)^{-1}_{11}&=& \frac{r_{+}(\la )}{E-E_+(\la )}+\frac{r_{-}(\la )}{E-E_-(\la )} \nonumber \\
(E-H)^{-1}_{22}&=& \frac{r_{-}(\la )}{E-E_+(\la )}+\frac{r_{+}(\la )}{E-E_-(\la )} \label{green}
\eea
while the corresponding poles of the off-diagonal terms have the
residues $\sqrt{r_{+}(\la )\cdot r_{-}(\la )}$.
With the insight from Eqs.(\ref{cxen},\ref{res},\ref{green}) we
qualitatively understand the shapes of Fig.3.  The strong
peak (left) originates from the near pole at $E_{+}(\la _{cr})$ with its strong
residue $r_{+}(\la )$. The right hand figure is produced by the
remote pole at $E_{-}(\la _r)$ with its strong residue while the small
peak on top is generated by the near pole with the very small
residue. The rather dramatic $\la -$dependence is due to
the proximity of the EPs; of course this is related to the
actual value of the coupling $\delta $ and will have a sensitive
dependence on $\delta $ at $\la =\la _{cr}$.
Note that, if the coupling in our two-dimensional model
is further increased, the narrow resonance is eventually
crossing the real energy axis and acquires a positive imaginary
part. It then can no longer be interpreted as a decaying state, that
is as a physical resonance. It may be interpreted, for example,
as an inversely populated level producing light amplification
generated by a radiation process in laser physics \cite{cao}.

\section{Conclusion}
Let us discuss a few cases where the phenomena
described in this paper have been encountered. In
a model for quantum transport \cite{kaplan}, transmission
increases drastically for a specific width
of some external barriers simulating the coupling strength of the
system with the environment; this parameter plays the role of our $\la $.
The authors stress that the complex eigenvalues of the effective Hamiltonian
are very important for understanding the transport properties of the
system. The analysis is based in part on findings of \cite{zel}
where the sensitive dependence upon the threshold energy
of the continuum coupling for loosely bound nuclei
has been demonstrated using an effective two by two non-Hermitian Hamiltonian.
In \cite{mosta} a specific frequency of the laser beam leads to
the occurrence of a 'resonance' pole with zero width; within its immediate
vicinity this parameter can be rewritten as our  parameter $\la $.
The quoted paper concentrates on specific solutions of the Schr\"odinger equation,
which are in fact the Gamow states (only outgoing waves as boundary conditions).
They correspond to the poles of the Green's function in our two-level model
(which has of course no continuous spectrum).
In all these cases, the extremely strong dependence on
the coupling of the pole positions
close to or upon the real axis, is the effect of near exceptional points.
The openness of the systems is crucial for all these phenomena.

In summary, the dramatic features of particular resonance behaviour
are explained by the square root singularities of spectrum and
eigenstates, especially the vanishing norm of the latter that gives
rise to the dramatic and distinctly different behaviour in each
channel. In a two channel problem, that was chosen
as a typical case, they invoke strong
dependence on the interaction strength  and significant
deviations from the usual patterns associated with isolated
or overlapping resonances.

\section*{Acknowledgements}
WDH is thankful for the hospitality which he received
from the Nuclear Theory Section of the Bogoliubov Laboratory, JINR during his
visit to Dubna. This work is partly supported by JINR-SA Agreement
on scientific collaboration, by Grant No. FIS2008-00781/FIS (Spain) and
RFBR Grants No. 08-02-00118 (Russia).


\begin{thebibliography}{}
\bibitem{foot}                    
                     here we do not consider bound states
                     occurring for negative energies for the same
                     boundary conditions; the continuous spectrum is treated
                     effectively using complex eigenvalues.
\end{thebibliography}
\end{document}